# On the Efficiency of Fast RSA Variants in Modern Mobile Phones

Klaus Hansen, Troels Larsen and Kim Olsen
Department of Computer Science
University of Copenhagen
Copenhagen, Denmark

*Abstract*—**Modern mobile phones are increasingly being used for more services that require modern security mechanisms such as the public-key cryptosystem RSA. It is, however, well-known that public-key cryptography demands considerable computing resources and that RSA encryption is much faster than RSA decryption. It is consequently an interesting question if RSA as a whole can be executed efficiently on modern mobile phones.**

**In this paper, we explore the efficiency on modern mobile phones of variants of the RSA cryptosystem, covering CRT, Multi-Prime RSA, Multi-Power RSA, Rebalanced RSA and R-Prime RSA by comparing the encryption and decryption time using a simple Java implementation and a typical RSA setup.**

## I. INTRODUCTION

Today, the virtually ubiquitous mobile phone is used for more complex services than just traditional voice calls and text messages. Many of these services require modern security mechanisms such as the Secure Sockets Layer (SSL) protocol. A number of these mechanisms, including SSL, make use of public-key cryptography.

It is, however, well-known that public-key cryptography demands considerable computing resources. On limited platforms such as the mobile phone, this problem is exacerbated and one may ask if it is possible at all to implement public-key cryptography efficiently on mobile phones, i.e. where the time needed to perform encryption or decryption is sufficiently small to avoid having a negative impact on the user experience while retaining the security of the cryptosystem.

In 2004, Großshadl and Tillich showed that both encryption and decryption using public-key cryptography based on elliptic curves could be executed efficiently on some mobile phones [7]. The authors also showed that encryption using the public-key cryptosystem RSA [1] was feasible but that decryption was not. To the best of our knowledge, no basic research results have been published showing that (the decryption of) RSA can be executed efficiently on mobile phones.

This paper investigates methods to optimize the execution of the decryption of RSA on modern mobile phones. We will do this by looking into various variants of the original RSA cryptosystem, thus taking an algorithmic approach rather than discussing hardware or software optimization schemes.

The paper is organized as follows. In section II, we introduce the original RSA cryptosystem and its variants (except Batch RSA, which has low relevance for our investigation), and in section III discuss our implementation of these cryptosystems in a specific mobile environment. In section IV, we show our experimental results of these cryptosystems on modern mobile phones and comment on the degree to which the variants improve the decryption time of the original RSA. Finally, in section V, we conclude the paper and take stock on the state of RSA as a whole on modern mobile phones, i.e. whether both encryption and decryption can be executed efficiently on modern mobile phones.

This paper is based on the results of a project carried out by the authors at the University of Copenhagen in the spring of 2008 [9].

## II. RSA AND ITS FAST VARIANTS

The original RSA cryptosystem was proposed in 1978 by Rivest, Shamir and Adelman [1] and consists of three parts:

- **Key generation:** Given an integer $n$, generate two different primes $p$ and $q$ of ($n/2$)-bit each and compute $N = pq$ and $\varphi(N) = (p-1)(q-1)$. Choose a random integer $1 < e < \varphi(N)$ such that $\gcd(e,\varphi(N)) = 1$. Next, compute the uniquely defined integer $1 < d < \varphi(N)$ satisfying $ed \equiv 1 \pmod{\varphi(N)}$. The public key is $<N,e>$ and the private key is $<N,d>$.

- **Encryption:** To encrypt a message $X$ with the public key $<N,e>$, transform the message $X$ to an integer $M$ in $\{0,\ldots,N-1\}$ and compute the ciphertext $C = M^e \bmod N$.

- **Decryption:** To decrypt the ciphertext $C$ with the private key $<N,d>$, compute $M = C^d \bmod N$ and employ the reverse transformation to obtain the message $X$ from $M$.[1]

---

[1] Note, we use the notation $a \bmod b$ to mean the remainder when $a$ is divided by $b$. The notation $a \equiv c \pmod{b}$ means that $a$ and $c$ result in the same remainder when divided by $b$, i.e. $a \bmod b = c \bmod b$.

Key generation is only performed occasionally so the efficiency of that part is less important than the two other parts, encryption and decryption. Their efficiency is determined by 1) the transformation of the message $X$ to the integer $M$, and back, and 2) the modular exponentiations $M^e$ mod $N$ and $C^d$ mod $N$. The transformations can be performed with standard algorithms, e.g. from the Public Key Cryptography Standards (PKCS) published by RSA Security [13]. This leaves the modular exponentiation as the single most important component of RSA in regards to its efficiency.

There exist a number of different methods for fast modular exponentiation [6]. In general, a modular exponentiation $a^b$ mod $c$ can be performed using $t$-1 modular squarings and $h$-1 modular multiplications where $t$ is the length of the binary representation of $b$ and $h$ is the number of 1's in the binary representation (i.e. its Hamming weight).

In practice, an often used method to perform modular exponentiations is the so-called repeated square-and-multiply algorithm [10]. In its binary version, the algorithm processes a single bit of the exponent $b$ at a time and on every iteration squares its intermediate result and multiplies it with $a$ if the current bit is set. Thus, the algorithm always performs $t$-1 modular squarings and at most $t$-1 modular multiplications. Since an upper bound on a single modular multiplication – and therefore also squaring – is $O(v^2)$, the repeated square-and-multiply algorithm has a running time of $O(tv^2)$ where $t$ is the bitlength of the exponent $b$ and $v$ is the bitlength of the modulus $c$.

Given this bound, the encryption exponent $e$ in the original RSA cryptosystem is typically chosen to be a small number, often $2^{16}$ + 1. There are two reasons for this: the relative short bitlength of $2^{16}$ + 1 will result in a small amount of modular squarings, and $2^{16}$ + 1 has only two 1's in its binary representation leading to the fewest possible modular multiplications for valid RSA encryption exponents, namely one [14]. Expressed informally, choosing $e$ this small almost effectively yields an encryption running time dependent only on the bitlength $n$ of the modulus $N$, i.e. $O(n^2)$. The structure of the decryption exponent $d$ cannot be tailored to fit the repeated square-and-multiply algorithm in the same way, but will often be long and consist of a random number of 1's in its binary representation. The worst case scenario is that $|d| \approx |N|$ yielding a running time of $O(n^3)$. This means that encryption is much faster than decryption in the original RSA.

A number of variants of the original RSA cryptosystem have been published over the years all seeking to improve the time-consuming decryption operation. We take a closer look at these variants in the following subsections.

*A. CRT RSA*

CRT RSA is the most commonly known RSA variant for speeding up decryption. It was first described by Couvreur and Quisquater in 1982 [5]. The idea behind CRT RSA is to split the costly decryption into two smaller and faster modular exponentiations using the Chinese Remainder Theorem, hence the acronym CRT RSA.

The Chinese Remainder Theorem tells us that a system of $r$ congruences

$x \equiv a_1 \pmod{n_1}, \ldots, x \equiv a_r \pmod{n_r}$,

where $n_1,\ldots,n_r$ are relatively prime integers and $a_1,\ldots,a_r$ ordinary integers has a unique solution modulo $N = n_1 n_2 \cdots n_r$. This solution can be written as

$x = (a_1 N_1 y_1 + \ldots + a_r N_r y_r) \bmod N$,

where $N_i = N/n_i$ and $y_i = N_i^{-1} \bmod n_i$ for $1 \leq i \leq r$ [15]. CRT RSA uses the Chinese Remainder Theorem the following way:

- **Key generation:** Generate $e$ and $d$ the same way as in the original RSA. Next, compute $d_p = d \bmod p$-1 and $d_q = d \bmod q$-1. The public key is $<N,e>$ and the private key $<p,q,d_p,d_q>$.

- **Encryption:** Encryption is the same as for the original RSA, $C = M^e \bmod N$.

- **Decryption:** Decryption is split into the following computations. First, compute $M_p = C^{d_p} \bmod p$ and $M_q = C^{d_q} \bmod q$. Then, using the Chinese Remainder Theorem, find $M$ as $M = (M_p q(q^{-1} \bmod p) + M_q p(p^{-1} \bmod q)) \bmod N$.

If we ignore the contribution from the sum function of the Chinese Remainder Theorem, decryption using CRT RSA requires two times $O((n/2)^3)$ since the bitlength of both the exponents and the moduli are $n/2$ and a single modular exponentiation has an upper bound of $O(tv^2)$ where $t$ is the bitlength of the exponent and $v$ is the bitlength of the modulus. Compared to the $O(n^3)$ decryption of the original RSA, CRT RSA improves decryption time with a factor

$$n^3 / (2 \cdot (n/2)^3) = 4.^2$$

*B. Multi-Prime RSA*

Multi-Prime RSA represents a natural generalisation of CRT RSA: By adding more primes to the generation of $N$, decryption can be split into an arbitrary number of smaller exponentiations instead of just two. This variant of the original RSA was first described by Collins *et al.* i 1997 [4]:

- **Key generation:** Given two integers $n$ and $r \geq 3$, generate $r$ different primes $p_1,\ldots,p_r$ each $(n/r)$-bit long. Set $N = \prod_{i=1}^{r} p_i$ and $\varphi(N) = \prod_{i=1}^{r} (p_i-1)$. Compute $e$ and $d$ as in the original RSA. Next, compute $d_i = d \bmod p_i$-1 for $1 \leq i \leq r$. The public key is $<N,e>$ and the private key $<p_1,\ldots,p_r,d_1,\ldots,d_r>$.

- **Encryption:** Encryption is the same as for the original RSA, $C = M^e \bmod N$.

- **Decryption:** Decryption is split into $r$ exponentiations, $M_i = C^{d_i} \bmod p_i$, for $1 \leq i \leq r$. Using the Chinese Remainder Theorem, $M$ is found as $M = (M_1 N_1 y_1 + \ldots + M_r N_r y_r) \bmod N$ where $N_i = N/p_i$ and $y_i = N_i^{-1} \bmod p_i$ for $1 \leq i \leq r$.

---

[2] This sort of big-o manipulation is formally not sound as big-o implies an arbitrary constant factor, but it does give a rough notion of the speed-up to expect in practice.

If we again ignore the contribution of the Chinese Remainder Theorem, decryption in Multi-Prime RSA requires $r$ times $O((n/r)^3)$. Compared to the $O(n^3)$ decryption of the original RSA, Multi-Prime RSA improves decryption time with a factor

$$n^3 / (r \cdot (n/r)^3) = r^2.$$

Obviously, the individual primes need to have a certain size to guard against factorisation attacks. This puts a natural limit to the size of $r$ and thus the actual improvement in decryption. Hinek suggest the following guidelines [8]:

| $n$ | 1024 | 2048 | 4096 | 8192 |
|---|---|---|---|---|
| max $r$ | 3 | 3 | 4 | 4 |

### C. Multi-Power RSA

Multi-Power RSA is a variant of Multi-Prime RSA. In Multi-Prime RSA, the modulus $N$ consists of $r$ different primes whereas the modulus in Multi-Power RSA has the structure $N = p^{r-1}q$ for $r \geq 3$. As shown below, this different structured modulus gives rise to a more efficient decryption than Multi-Prime RSA. Multi-Power RSA was first described by Takagi in 1998 [16]:

- **Key generation:** Given two integers $n$ and $r \geq 3$, generate two different primes $p$ and $q$ each $(n/r)$-bit long. Set $N = p^{r-1}q$. Compute $e$ as in the original RSA and $d$ satisfying $ed \equiv 1 \pmod{(p-1)(q-1)}$. Finally, compute $d_p = d \bmod p-1$ og $d_q = d \bmod q-1$. The public key is $<N,e>$ and the private key $<p,q,d_p,d_q>$.

- **Encryption:** Encryption is the same as for the original RSA, $C = M^e \bmod N$.

- **Decryption:** Conceptually, decryption is the same as CRT RSA. First, compute $M_q = C_q^{dq} \bmod q$ where $C_q = C \bmod q$ and $M_p = C_p^{dp} \bmod p^{r-1}$ where $C_p = C \bmod p^{r-1}$. Next, apply the Chinese Remainder Theorem to find $M$.

Using Hensel Lifting, $M_p$ can be computed using only one modular exponentiation modulu $p$ instead of modulu $p^{r-1}$, and some extra arithmetic. Ignoring the contribution from the Chinese Remainder Theorem and the extra arithmetic from the Hensel Lifting, this means that decryption in Multi-Power RSA requires two times $O((n/r)^3)$. Compared to the $O(n^3)$ decryption of the original RSA, Multi-Power RSA improves decryption time with a factor

$$n^3 / (2 \cdot (n/r)^3) = r^3/2.$$

With respect to the security of Multi-Power RSA, the same guidelines as for Multi-Prime RSA apply limiting the size of $r$ and thus the actual improvement in practice. The special Lattice Factoring Method designed to factor integers with the structure $N = p^rq$ is not a security issue for Multi-Power RSA as it cannot be efficiently applied for usual sizes of $N$ in RSA [3].

### D. Rebalanced RSA

Basically, in the original RSA, encryption is more efficient than decryption because $e$ is small and $d$ is large. A straightforward way to optimise decryption would be to "switch" the exponents, i.e. make $e$ large and $d$ small. However, this is not to be recommended. Small values of $d$ open up for Wiener's Low Decryption Exponent Attack [2]. Instead Wiener proposed in 1990 the variant Rebalanced RSA [17] that retains the size of $d$ but makes $d_p = d \bmod p-1$ and $d_q = d \bmod q-1$ small (at the expense of a larger $e$):

- **Key generation:** Given integers $n$ and $w$, generate two different primes $p$ and $q$ each $(n/2)$-bit long such that $\gcd(p-1,q-1) = 2$. Set $N = pq$ and $\varphi(N) = (p-1)(q-1)$. Compute two $w$-bit integers $d_p$ and $d_q$ satisfying $\gcd(d_p,p-1) = \gcd(d_q,q-1) = 1$ and $d_p \equiv d_q \pmod{2}$. Find a $d$ such that $d = d_p \bmod p-1$ and $d = d_q \bmod q-1$. Compute $e = d^{-1} \bmod \varphi(N)$. The public key is $<N,e>$ and the private key $<p,q,d_p,d_q>$.

- **Encryption:** Encryption is the same as for the original RSA, $C = M^e \bmod N$, but with a much larger $e$ (on the order of $N$).

- **Decryption:** Decryption is the same as for CRT RSA but with smaller $d_p$ and $d_q$ – each $w$-bit long in Rebalanced RSA versus $(n/2)$-bit in CRT RSA. Usually, $w \geq 160$ and $n/2 \geq 512$.

If we again ignore the contribution of the Chinese Remainder Theorem, decryption in Rebalanced RSA requires two times $O(w(n/2)^2)$. Compared to the $O(n^3)$ decryption of the original RSA, Rebalanced RSA improves decryption time with a factor

$$n^3 / (2w \cdot (n/2)^2) = 2n/w.$$

With respect to the security of Rebalanced RSA, it is recommended to set $w \geq 160$ thereby limiting the actual improvement of decryption in practice. Note also that the speed-up in decryption comes at the cost of a much slower encryption. In the original RSA, $e$ is small – typically 16 bit or less – whereas in Rebalanced RSA, $e$ is on the order of $N$. This means that encryption in Rebalanced RSA is as slow as decryption in the original RSA.

### E. R-Prime RSA

In Rebalanced RSA, decryption is the same as in CRT RSA. Since Multi-Prime RSA is a generalisation of CRT RSA, why not generalise Rebalanced RSA to use Multi-Prime RSA in its decryption as well. This is the idea behind R-Prime RSA, first described by Paixão in 2003 [12]:

- **Key generation:** Given $n$ and $w$, generate $r \geq 3$ different primes $p_1,\ldots,p_r$ each $(n/r)$-bit long such that $\gcd(p_1-1,\ldots,p_r-1) = 2$. Set $N = \prod_{i=1}^r p_i$ and $\varphi(N) = \prod_{i=1}^r (p_i-1)$. Compute $r$ $w$-bit integers $d_{p1},\ldots,d_{pr}$ satisfying $\gcd(d_{p1},p_1-1) = \ldots = \gcd(d_{pr},p_r-1) = 1$ and $d_{p1} \equiv \ldots \equiv d_{pr}$ (mod 2). Find a $d$ such that $d = d_{p1}$ mod $p_1$-1,…, $d = d_{pr}$ mod $p_r$-1. Compute $e = d^{-1}$ mod $\varphi(N)$. The public key is $<N,e>$ and the private key $<p_1,…,p_r,d_{p1},…,d_{pr}>$.

- **Encryption:** Encryption is the same as for the original RSA, $C = M^e$ mod $N$, but with a much larger $e$ (as was the case with Rebalanced RSA).

- **Decryption:** Decryption is the same as for Multi-Prime RSA. That is, decryption is split into $r$ modular exponentiations $M_i = C^{dpi}$ mod $p_i$ for $1 \leq i \leq r$ after which the Chinese Remainder Theorem is applied. The difference lies in the length of $d_{pi}$ (denoted $d_i$ in Multi-Prime RSA). In R-Prime RSA, these values are $w$-bit each. In Multi-Prime RSA, they are $n/r$ each.

This means that the decryption in R-Prime RSA requires $r$ times $O(w(n/r)^2)$ (ignoring the Chinese Remainder Theorem). Compared to the $O(n^3)$ decryption of the original RSA, R-Prime RSA improves decryption time with a factor

$$n^3 / (rw \cdot (n/r)^2) = nr/w.$$

R-Prime RSA inherits the same security considerations as Rebalanced RSA and Multi-Prime RSA, i.e. it is recommended to set $w \geq 160$ and limit the value of $r$ with the respect to $n$ as listed earlier. As in Rebalanced RSA, the speed-up in decryption means a much slower encryption.

Since Multi-Power RSA is faster than Multi-Prime RSA, why not use Multi-Power RSA in R-Prime RSA? The reason lies in the technique Hensel Lifting that is used in decryption in Multi-Power RSA. Hensel Lifting makes use of a number of exponentiations modulo $e$ which is not a problem in Multi-Power RSA since $e$ is small. But in R-Prime RSA, $e$ is on the order of $N$ making Hensel Lifting (and consequently, Multi-Power RSA) disproportionately expensive.

*F. Comparison*

The following table compares the decryption time of the original RSA to all its variants (excluding Batch RSA). The table shows the complexity of decryption for each variant, their general theoretical improvement and their approximated improvement in practice when $n = 1024$, and $r = 3$ and $w = 160$ where applicable:

| Variant | Complexity | Theo. | Appr. |
|---|---|---|---|
| Original | $O(n^3)$ | 1.0 | 1.0 |
| CRT | $2 \cdot O((n/2)^3)$ | 4.0 | 4.0 |
| Multi-Prime | $r \cdot O((n/r)^3)$ | $r^2$ | 9.0 |
| Multi-Power | $2 \cdot O((n/r)3)$ | $r^2/2$ | 13.5 |
| Rebalanced | $2 \cdot O(w(n/2)^2)$ | $2n/w$ | 12.8 |
| R-Prime | $r \cdot O(w(n/r)^2)$ | $nr/w$ | 19.2 |

No single variant outperforms all the others. Which variant is best is a question of usage scenario. If both encryption and decryption is needed, then Multi-Prime or Multi-Power RSA should be best. If only encryption is needed, there is no need to upgrade from the original RSA. And if only decryption is needed, then R-Prime RSA promises the most improvement.

III. IMPLEMENTATION

We have implemented the original RSA and all its variants (except Batch RSA) to demonstrate their actual improvement in practice on modern mobile phones.

As most mobile phones support Java's runtime environment, our implementation is written in Java's mobile platform Java 2 Micro Edition (J2ME). J2ME contains a subset of the Java Standard Edition (JSE). The Connected Limited Device Configuration (CLDC) and the Mobile Information Device Profile (MIDP) define the available APIs. We use the CLDC 1.1 and MIDP 2.1. These contain no general cryptographic API or a support for multi-precision computations like the BigInteger class. Therefore, we base the mathematical operations such as modular exponentiations on the BigInteger class from the bouncycastle.org [11].

The lack of built-in cryptographic APIs has the effect that each java application that require the use of cryptographic funcitons has to be bundled with code that provides such features. This leads to larger code sizes (in our case around 100 KB) and prevents code sharing.

Finally, note that we do not implement any textual transformations but only the core encryption and decryption of $M$ and $C$, respectively.

IV. EXPERIMENTAL RESULTS

Our implementation of the original RSA and its variants is tested on a HTC Touch Dual from 2007 with a Qualcomm MSM 7200 400MHz processor and 128MB SDRAM. We test encryption and decryption of the original RSA and decryption for all the variants. We set $n = 1024$, and $r = 3$ and $w = 160$ where applicable. $M$ is the same for all tests.

The following table summarizes the results of our experiment by listing the average of twenty decryptions for each cryptosystem (and encryptions for the original RSA), the actual improvement in decryption and the approximated improvement in decryption (all figures are in ms):

| Variant | Enc. | Dec. | Actual | Appr. |
|---|---|---|---|---|
| Original | 29 | 2098 | 1.0 | 1.0 |
| CRT | - | 558 | 3.8 | 4.0 |
| Multi-Prime | - | 283 | 7.4 | 9.0 |
| Multi-Power | - | 210 | 10.0 | 13.5 |
| Rebalanced | - | 187 | 11.2 | 12.8 |
| R-Prime | - | 151 | 13.9 | 19.2 |

As expected, the actual improvement for each variant is less than the approximated improvement due to overhead from the Chinese Remainder Theorem and other arithmetic. This is most outspoken for Multi-Power and R-Prime making Rebalanced faster than Multi-Power while R-Prime retains its position as the fastest decryptor. As evident, the approximated improvements are useful as rough guidelines.

It is interesting to note the relative improvement from CRT to Multi-Prime is bigger than the relative improvement from Rebalanced to R-Prime. An closer inspection of the variants reveals that this is due to the fact that the generalization from CRT to Multi-Prime results in both smaller exponents *and* moduli while the generalization from Rebalanced to R-Prime only leads to smaller moduli as the exponents are fixed at $w = 160$.

All in all, our experimental results show that the actual decryption time of each variant is well below one second – and verify that the actual encryption time of the original RSA is well below that of all the decryption times.

## V. CONCLUSION

We have implemented the central part of various variants of the RSA cryptosystem: CRT RSA, Multi-Prime RSA, Multi-Power RSA, Rebalanced RSA and R-Prime RSA and run a test on typical data that shows that they improve the decryption time of the original RSA considerably, achieving actual decryption times well below one second.

Consequently, we are able to assert that both RSA encryption and decryption can be executed efficiently on a modern mobile phone.